\documentclass[prl,twocolumn,superscriptaddress,nofootinbib]{revtex4-2}
\usepackage{epsfig}
\usepackage{graphicx}
\usepackage{palatino}
\usepackage[english]{babel}
\usepackage{hyphenat}
\usepackage{amsmath}
\usepackage{amssymb}
\usepackage{slashed}
\usepackage{epstopdf}
\usepackage{xcolor}
\usepackage{booktabs}
\usepackage{lipsum, babel}
\usepackage[toc,page]{appendix}
\definecolor{lcolor}{rgb}{0.,0.0,0.}
\definecolor{citcolor}{rgb}{0,0.,0.5}
\usepackage[breaklinks,colorlinks,urlcolor=blue,citecolor=blue,linkcolor=blue]{hyperref}
\usepackage{multirow}
\usepackage{ltablex}
\usepackage{bbm}

\newcommand{\beq}{\begin{equation}}
\newcommand{\eeq}{\end{equation}}
\newcommand{\bea}{\begin{eqnarray}}
\newcommand{\eea}{\end{eqnarray}}

\def\dd{{\rm d}}

\newcommand{\bem}{\begin{multline}}
\newcommand{\eem}{\end{multline}}
\newcommand{\beg}{\begin{gather}}
\newcommand{\eeg}{\end{gather}}

\newcommand{\ben}{\begin{eqnarray*}}
\newcommand{\een}{\end{eqnarray*}}

\newcommand{\bal}{\begin{align}}
\newcommand{\eal}{\begin{align}}

\def\Tr{\text{Tr}}

\def\and{\quad\text{and}\quad}

\def\0{{\boldsymbol 0}}

\def\p{{\boldsymbol p}}

\def\k{{\boldsymbol k}}

\def\b{{\boldsymbol b}}

\def\R{{\boldsymbol R}}

\newcommand{\vect}[1]{\boldsymbol{#1}_{\perp}}
\newcommand{\kt}{\vect{k}}
\newcommand{\ktone}{\vect{k_{1}}}

\newcommand{\pt}{\vect{p}}
\newcommand{\Pt}{\vect{P}}
\newcommand{\pgt}{\boldsymbol{p}_{g\perp}}
\newcommand{\rt}{\vect{r}}
\newcommand{\qt}{\vect{q}}

\newcommand{\bt}{\vect{b}}
\newcommand{\xt}{\vect{x}}

\newcommand{\Rt}{\vect{R}}
\newcommand{\lt}{\vect{l}}
\newcommand{\der}{\mathrm{d}}

\begin{document}
\title{Accessing subnuclear fluctuations and saturation with multiplicity dependent\\ $J/\psi$ production in $p+p$ and $p+Pb$ collisions}
\author{Farid Salazar}
\email[]{salazar@physics.ucla.edu}
\affiliation{Department of Physics and Astronomy, University of California, Los Angeles, CA 90095, USA}
\affiliation{Mani L. Bhaumik Institute for Theoretical Physics, University of California, Los Angeles, CA 90095, USA}
\affiliation{Nuclear Science Division, Lawrence Berkeley National Laboratory, Berkeley, CA 94720, USA}
\affiliation{Physics Department, University of California, Berkeley, CA 94720, USA}
\author{Björn Schenke}
\email[]{bschenke@bnl.gov}
\affiliation{Physics Department, Brookhaven National Laboratory, Upton, NY 11973, USA}
\author{Alba Soto-Ontoso}
\email[]{alba.soto@ipht.fr}
\affiliation{Universit\'e Paris-Saclay, CNRS, CEA, Institut de physique th\'eorique, 91191, Gif-sur-Yvette, France}

\begin{abstract}
We study the production of $J/\psi$ vector mesons as a function of charged hadron multiplicity in $p+p$ and $p+Pb$ collisions at LHC energies. We employ the color glass condensate framework, using running coupling Balitsky-Kovchegov evolved dipole amplitudes, to compute gluon and $c\bar{c}$-pair production. We use fragmentation functions to obtain charged hadrons, and explore two different hadronization schemes for the $J/\psi$: non-relativistic quantum chromodynamics and the improved color evaporation model. In our framework, event-by-event multiplicity fluctuations of both hadrons and $J/\psi$ are driven by geometric and saturation scale normalization fluctuations. Studying the correlation between $J/\psi$ and hadron multiplicity, we show that the characteristic difference  between forward and backward rapidity in $p+Pb$ collisions is a result of different degrees of saturation probed at different rapidities. 
We demonstrate that experimental data on heavy-flavor production as a function of event activity provide stringent constraints on the fluctuating proton structure.
\end{abstract}

\maketitle

\section{Introduction}
Observing signals of gluon saturation in hadronic collisions as well as deeply inelastic scattering has been a long standing goal in the high energy nuclear physics community. While many observables are compatible with saturation model results or even more sophisticated color glass condensate (CGC) effective field theory calculations, see \cite{Iancu:2003xm,Gelis:2010nm,Kovchegov:2012mbw} and references therein, so far there has been no clear evidence that we have reached the kinematic regime where gluon saturation is present and strongly affects the particle production \cite{Albacete:2014fwa,Blaizot:2016qgz,Morreale:2021pnn}.

The study of rapidity dependent vector meson production in hadronic collisions, in combination with charged hadron production, is well suited to test saturation effects at a semi-hard scale of a few GeV \cite{Watanabe:2016ert}. In a nutshell, heavy vector mesons, whose mass is of the order of magnitude of the saturation scale, will be less sensitive to saturation effects than lighter charged hadrons at similar transverse momenta and rapidities. Because the saturation scale depends on (i) the mass number of the nuclei, (ii) the collision energy, and (iii) produced particles' four-momenta, asymmetric collision systems offer the possibility of exploring potential effects of saturation systematically in this observable.

The inclusive production of vector mesons as a function of the charged hadron multiplicity or event activity has been measured at the Large Hadron Collider (LHC) in both $p+p$ \cite{ALICE:2012pet,CMS:2013jsu,Adam:2015ota,ALICE:2020msa} and $p+Pb$  \cite{CMS:2013jsu,ALICE:2017wet,ALICE:2020eji} collisions at various center of mass energies. In $p+p$ and $p+Pb$ collisions at mid-rapidity, one finds a faster than linear growth of the vector meson yield with the charged hadron multiplicity. As a function of mid-rapidity charged hadron multiplicity, the vector meson yield rises faster than linearly in the $Pb$-going direction and more slowly than linear in the $p$-going direction. 

A variety of model calculations have been employed to compute this observable. In $p+p$ collisions, the faster than linear growth is reproduced by all models \cite{Sjostrand:2014zea,Adam:2015ota,Weber:2018ddv,Ferreiro:2012fb,Werner:2013tya,Werner:2010aa,Kopeliovich:2013yfa,Kopeliovich:2019phc} compared in \cite{ALICE:2020msa}, yet the effect is attributed to different underlying physics mechanisms. 

Within the color glass condensate framework, which we employ in this work, a faster than linear growth is explained by the fact that the saturation scale depends on the longitudinal momentum fraction of probed partons, $x$, and that higher mass vector mesons are sensitive to larger $x$ values than the lower mass charged hadrons. As saturation effects increase with increasing $Q_s$ (decreasing $x$), one expects a relative suppression of the light mass particles compared to the heavier mass ones, leading to the observed faster than linear growth. In addition, the higher mass vector meson probes smaller transverse length scales, where saturation effects are less dominant.  We will explain this picture in more detail and discuss the uncertainties related to e.g. hadronization below. 

The rapidity dependence of the $J/\psi$ production as a function of the midrapidity charged hadron yield in $p+Pb$ collisions has been well reproduced within the EPOS3 model without hydrodynamics \cite{ALICE:2020eji}. Studies within the CGC have compared to the $p+Pb$ result at mid-rapidity \cite{Ma:2018bax}. In this study, the typical $Q_s$ values in the lead and proton were varied by hand in order to generate high multiplicity configurations. However, there has not been a detailed CGC calculation with realistic multiplicity fluctuations that could explain the particle production at forward and backward rapidities in $p+Pb$ collisions. 

In this work we aim to go beyond previous calculations in the CGC framework by including fluctuations of the geometry and local densities in projectile and target, for both $p+p$ and $p+Pb$ collisions. We compute gluon production using $k_{T}$-factorized cross sections \cite{Kovchegov:2001sc,Blaizot:2004wu} and hadronize using KKP fragmentation functions \cite{Kniehl:2000fe}. To compute the production of $J/\psi$ mesons we make a large $N_c$ approximation \cite{Fujii:2006ab}, and use either non-relativistic quantum chromodynamics (NRQCD) \cite{Bodwin:1994jh,Kang:2013hta,Ma:2014mri} or the improved color evaporation model (ICEM) \cite{Ma:2016exq}. Further, we employ small-$x$ evolution based on the running coupling Balitsky-Kovchegov (BK) equations \cite{Balitsky:1995ub,Kovchegov:1996ty}. 
We focus on the correlation between $J/\psi$ and charged hadron multiplicities, as well as the average transverse momentum of the $J/\psi$ as a function of event activity. However, the model parameters are tuned to a wider set of observables, including transverse momentum and rapidity distributions of charged hadrons and $J/\psi$, which will be presented in a separate publication \cite{long:paper}. 

\section{Theoretical Model}
In the CGC formalism, a highly energetic parton in color representation $\rm R$ (either fundamental $\rm{R}=F$ for a quark, or adjoint $\rm{R}=\rm{Adj}$ for a gluon) acquires a transverse momentum $\kt$\footnote{In this work, we denote two dimensional transverse vectors and their magnitude as $\kt$ and $k_\perp$ respectively. We use the short-hand notation for the transverse integration $\int_{\kt} = \int \der^2 \kt$.} via propagation through the dense target color field. This process is described in terms of a two point correlator of Wilson lines, $V_{\mathrm{R}}$:
\begin{align}
    \!\!\!\!\widetilde{\mathcal{S}}_{\rm{R}}(x;\kt;\Rt) &\!= \!\!\int_{\rt} \!\!\!\!\! e^{\!-i \kt \!\cdot \rt} \! S_{\rm{R}} \! \left(\! x; \Rt\!+\!\frac{\rt}{2}, \Rt \!-\! \frac{\rt}{2}\!\right)\!,\!\!
    \label{eq:2pointWilson}
\end{align}
with
\begin{align}
    &S_{\rm{R}}(x;\xt,\xt') =\frac{1}{d_{\mathrm{R}}}\left \langle \Tr\left[ V_{\mathrm{R}}(\xt) V_{\mathrm{R}}^\dagger(\xt') \right] \right \rangle_{x} \,,
    \label{eq:2pointWilson-coordinate}
\end{align}
where $d_{\mathrm{R}}$ denotes the dimension of the representation.
The impact parameter of the parton relative to the center of the target is given by $\Rt$. In addition to the $x$ (or rapidity) dependence, in this work the CGC ensemble average $\left \langle \dots \right \rangle_{x}$ also depends on the underlying model for the nucleus, which will be described in more detail in a subsequent section.
This Wilson line correlator is the key ingredient of our computation, and will enter cross sections for both  charged hadron and $J/\psi$ production.

\subsection{Charged hadron production}
In the high energy limit, we assume that the gluon channel dominates charged hadron production. We begin with gluon production in proton-nucleus ($p+A$) collisions\footnote{Throughout this manuscript, in our definition of proton-nucleus collisions we also include proton-proton collisions.} in the CGC, and then discuss the hadronization mechanism using KKP fragmentation functions.

\begin{figure}[ht]
    \centering
    \includegraphics[scale=0.28]{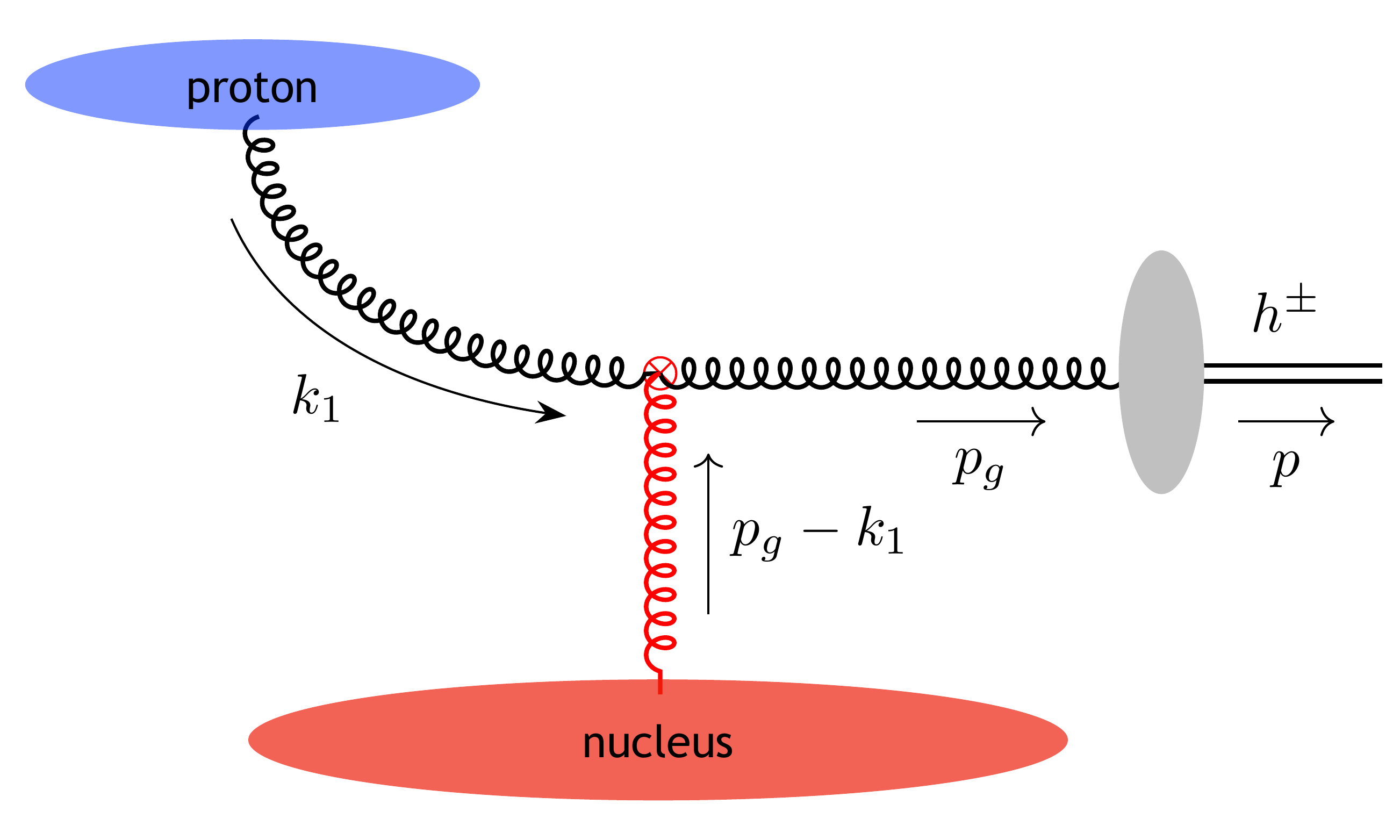}
    \caption{Leading order diagram for charged hadron production in proton-nucleus collisions within the CGC. The vertical gluon line represents the multiple scattering between a projectile gluon and the field of the nucleus. The gray blob represents the hadronization of gluons to hadrons.}
    \label{fig:gluonprod}
\end{figure}

In the CGC and within $k_{T}$-factorization \cite{Kovchegov:2001sc,Blaizot:2004wu}, the gluon yield (at fixed impact parameter $\bt$) differential in transverse momentum $\pgt$ and rapidity $y_g$ reads
\begin{align}
    \label{eq:g_xsection}
    \frac{\der N_g (\bt)}{\der^2 \pgt \der y_g} = &\frac{\alpha_s}{(\sqrt{2}\pi)^6 C_F} \frac{1}{\pgt^2} \int_{\ktone,\Rt} \!\!\!\!\!\! \phi^{p}(x_{p};\ktone;\Rt) \nonumber \\
    &\times\phi^{A}(x_{A};\pgt-\ktone;\Rt-\bt) \,,
\end{align}
where $A=p,Pb$ for $p+p$ and $p+Pb$ collisions, respectively. In Eq.\,\eqref{eq:g_xsection}, 
$\ktone$ is the transverse momentum of the incoming gluon that exchanges a transverse momentum $\pgt -\ktone$ with the nucleus (see Fig.\,\ref{fig:gluonprod}). For any nucleus, the (impact parameter dependent) unintegrated gluon distribution \cite{Kovchegov:2001sc,Blaizot:2004wu}
\begin{align}
    \phi(x;\kt;\Rt) &= \frac{\kt^2 C_F}{2\alpha_s} \widetilde{\mathcal{S}}_{\rm{Adj}}(x;\kt;\Rt) \,,
    \label{eq:UGD-alltwist}
\end{align}
is obtained using Eq.\,\eqref{eq:2pointWilson}. The longitudinal momentum fractions $x$ follow from kinematics and momentum conservation: $x_{p/A}= p_{g\perp}\exp(\pm y_g)/\sqrt{s}$. The strong coupling constant $\alpha_s$ is treated as a free parameter and $C_F=(N_c^2-1)/(2 N_c)$, with the number of colors fixed to $N_c=3$. 

In the dilute-dense approximation, the unintegrated gluon distribution of the proton $\phi^p$ should be obtained from the leading twist expansion of Eq.\,\eqref{eq:2pointWilson}. In this work, we follow the ansatz employed in \cite{Blaizot:2004wu,Ma:2014mri,Ma:2018bax}, which attempts to resum a subset of saturation corrections in the proton. The importance of saturation corrections has been quantified in \cite{Schlichting:2019bvy}, and a systematic program to semi-analytically include these corrections is underway \cite{Li:2021zmf,Li:2021yiv,Li:2021ntt}. Another possibility would be to numerically compute gluon production in the dense-dense limit \cite{Krasnitz:2000gz,Krasnitz:2001qu,Lappi:2003bi}.  

To obtain the charged hadron pseudo-rapidity density $\dd N_{\rm ch}/\dd\eta$ from the gluon spectrum, we employ the next-to-leading order (NLO) KKP fragmentation functions $D_h(z)$ \cite{Kniehl:2000fe} with fragmentation scale $\mu=p_{\perp}$ and $z=p_{\perp}/p_{g\perp}$, and integrate over the hadron transverse momentum $\p_\perp$:
\begin{align}
    \frac{\der N_{\rm ch}(\bt)}{\der \eta} = \!\! \int_{\pt} \!\! \int_{z_{\rm min}}^1 \hspace{-0.3cm} \der z \frac{D_h(z)}{z^2} \mathcal J_{y\rightarrow \eta} 
    \left. \frac{\der N_{g}(\bt)}{\der^2 \pgt \der y_g } \right|_{\pgt=\pt/z} \,,
    \label{eq:hadrons}
\end{align}
where $z_{\rm min}$ is determined from the condition that both $x_{p}$ and $x_{A}$ need to be smaller than 1.
We assume that the gluon rapidity equals that of the hadrons, $y_g=y$, with
\begin{equation}
    y=\frac{1}{2}\ln\displaystyle\frac{\sqrt{p^2_{\perp}\cosh^2{\eta}+m_h^2}+p_{\perp}\sinh{\eta}}{\sqrt{p^2_{\perp}\cosh^2{\eta}+m_h^2}-p_{\perp}\sinh{\eta}}\,,
\end{equation}
where $m_h=0.2\,{\rm GeV}$ is the effective hadron mass. Finally, the Jacobian 
$\mathcal J_{y\rightarrow \eta} =p_{\perp}\cosh{\eta}/\sqrt{p^2_{\perp}\cosh^2{\eta}+m_h^2}$
is required to convert rapidity $y$ to pseudo-rapidity $\eta$. Note that in $p+Pb$ collisions, we shift the gluon distribution to the laboratory frame by $\Delta y = 0.465$ before performing the fragmentation.

An infrared regulator on the $\p_\perp$ integral in Eq.\,\eqref{eq:hadrons} is required. We choose to impose it on the hadron momentum by replacing the expression $1/(p_{g\perp}^2 z^2)$, that appears in Eq.\,\eqref{eq:hadrons} after inserting Eq.\,\eqref{eq:g_xsection}, by
$1/(p_{g\perp}^2 z^2+m_{\rm IR}^2)$. By doing this instead of regulating the gluon spectrum, we simultaneously compensate for the overestimation of the hadron spectrum at $p_\perp<1\,{\rm GeV}$, which is a consequence of using a fragmentation function that uses perturbative evolution equations \cite{Kniehl:2000fe}.

\subsection{Vector meson production}
The direct\footnote{As done in previous works \cite{Ma:2014mri}, we neither consider feed-down from  $\psi'$ and $\chi_c$, which is part of the prompt $J/\psi$ production, nor decays of b-hadrons, which contribute to inclusive production.} production of quarkonium in proton-nucleus collisions occurs when a gluon from the proton fluctuates, either before or after multiply scattering off the field of the nucleus, into a pair of heavy quarks, which eventually forms a bound state of quarkonium. We employ two mechanisms for the formation of the quarkonium state: the improved color evaporation model and non-relativistic QCD.

\emph{Improved Color Evaporation model:} The improved color evaporation model describes the formation of $J/\psi$ with the compact formula presented in  \cite{Ma:2016exq}
\begin{align}
\label{eq:ICEM}
    \frac{\der N_{J/\psi}(\bt)}{\der^2 \Pt \der Y} = F \! \int_{m^2_{J/\psi}}^{4m^2_D} \!\!\!\!\!\!\!\! \der M^2 \frac{M^2}{m_{J/\psi}^2} \!\!\left.\frac{\der N_{c\bar c}(\b_\perp)}{\der M^2 \der^2 \Pt \der Y}\right|_{\Pt'}\,,
\end{align}
where $F$ represents the probability for a $c\bar c$ pair to form a $J/\psi$, and $M^2$ and $\Pt$ are the invariant mass squared and transverse momentum of the $c\bar{c}$ pair, respectively. In the ICEM, the $c\bar{c}$ yield is evaluated at $\Pt'= (M/m_{J/\psi}) \Pt$. We use $F=0.1$, which is approximately $40\%$ bigger than the one used in \cite{Ducloue:2015gfa}. However, this quantity only enters as an overall normalisation in the calculation and we tuned it to reproduce the $J/\psi$ spectrum. It does not directly affect the observables studied in this work.

The inclusive production yield of a $c\bar{c}$ pair in proton-nucleus collisions in the CGC is given by \cite{Fujii:2006ab,Fujii:2013gxa,Fujii:2013yja}
\begin{widetext}
\begin{align}
    \label{eq:cgc-ccbar}
    \frac{\der N_{c\bar c}(\b_\perp)}{\der^2\pt \der^2 \qt \der y_c \der y_{\bar c}} = &\frac{\alpha_s N_c^2}{2(2\pi)^{10}(N_c^2-1)} \int_{\ktone;\kt;\Rt} \!\!\! \frac{\phi^{p}(x_p;\ktone;\Rt)}{\ktone^2} \widetilde{\mathcal{S}}_{\rm{F}}^{A}(x_A;\k_{\perp};\R_\perp-\b_\perp)   \nonumber \\
    \times & \widetilde{\mathcal{S}}_{\rm F}^{A}(x_A;\pt + \qt -\ktone - \kt;\R_\perp-\b_\perp) \mathcal{H}(\pt,\qt,\ktone,\pt + \qt -\ktone - \kt) + \mathcal{O}\left(1/N_c\right) \,.
\end{align}
\end{widetext}

\begin{figure}[ht]
    \centering
    \includegraphics[scale=0.28]{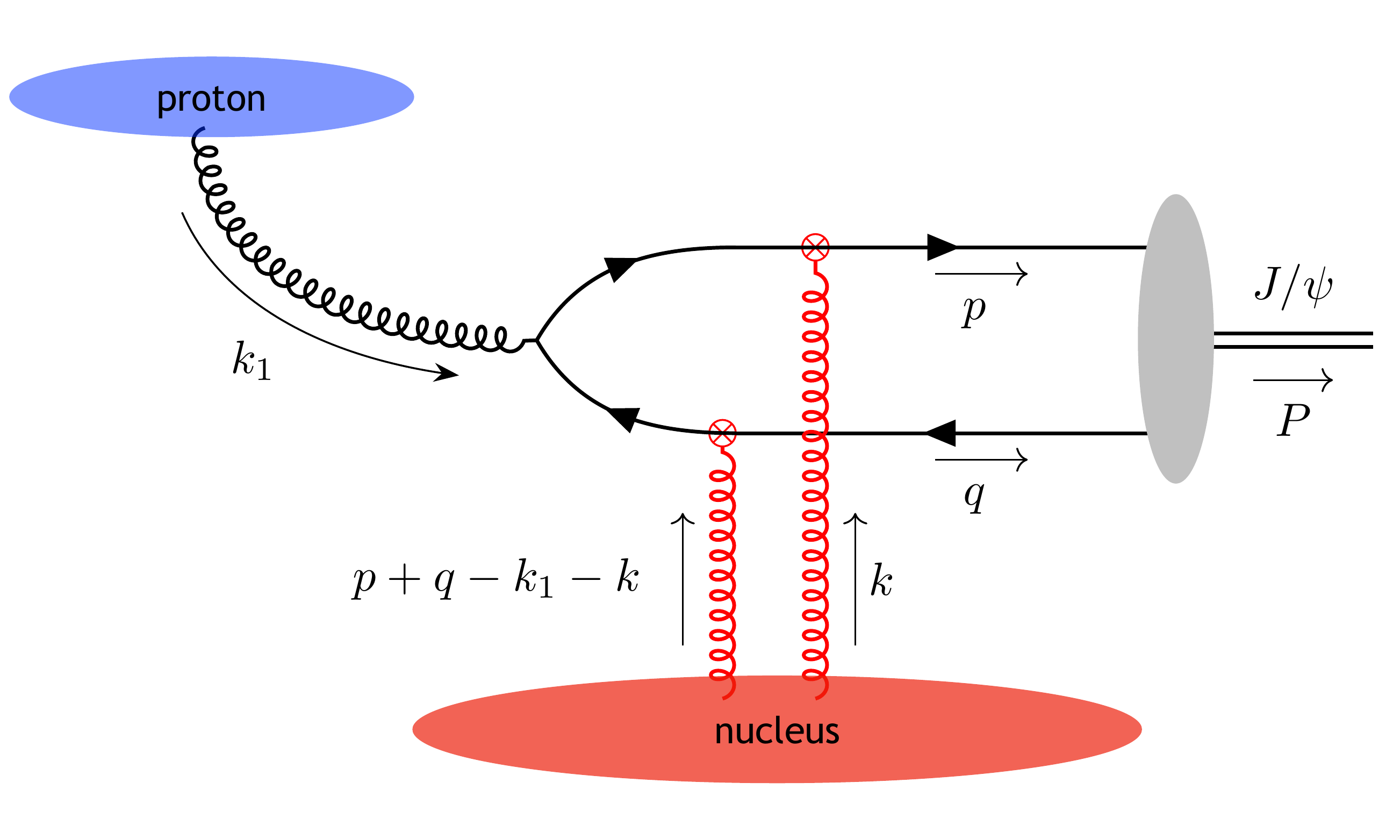}
    \caption{One of the two leading order diagrams for $J/\psi$ production in $p+A$ collisions within the CGC. The two gluons from the target represent the multiple scattering between the dipole and the field of the nucleus. Hadronization is represented by the gray blob. The other diagram has the vertical gluon line attached to the incoming gluon instead.}
    \label{fig:ccbarprod}
\end{figure}

We have only kept the leading $N_c$ contribution, which means that only the color octet channel contributes, and the color singlet is neglected. In Eq.\,\eqref{eq:cgc-ccbar}, $\Rt$ and $\Rt-\bt$ are the transverse location of the gluon relative to the center of the proton and center of the nucleus respectively\footnote{Due to the mass of the charm quark, the $c\bar{c}$ dipole is small, thus we assume that the location of the produced $c$ and $\bar{c}$ is approximately that of the gluon.}, $\ktone$ is the transverse momentum of the incoming gluon, and $\kt$ is the transverse momentum transferred from the nucleus to the quark. The transverse momentum exchanged between the anti-quark and the nucleus is fixed by momentum conservation (see Fig.\,\ref{fig:ccbarprod}). The hard matrix element, $\mathcal H$, takes into account the splitting of the gluon into the $c\bar c$ pair either before or after the interaction with the nucleus, together with the interference term. The explicit expression of the hard factor is given in~\cite{Fujii:2013gxa,Ma:2018bax}. As in the charged hadron production subsection, we will employ the ansatz in \cite{Blaizot:2004wu,Ma:2014mri,Ma:2018bax}
and compute $\phi^p(x_p;\ktone;\Rt)$ from Eq.\,\eqref{eq:UGD-alltwist} instead of the leading twist expansion. 

To convert Eq.\,\eqref{eq:cgc-ccbar} into the form required by Eq.\,\eqref{eq:ICEM}, we apply the transformation of coordinates \cite{Fujii:2006ab}:
\begin{align}
    \label{eq:cgc-ccbar-pair}
    \frac{\der N_{c\bar c}(\bt)}{\der M^2 \der^2 \Pt \der Y} = \int_0^{\sqrt{\frac{M^2}{4}-m_c^2}}\hspace{-0.5cm} \der \tilde q \int_0^{2\pi} \hspace{-0.3cm} \der \phi \mathcal{J}  \frac{\der N_{c\bar c}(\bt)}{\der^2 \pt \der^2 \qt \der y_p \der y_{q}} \,,
\end{align}
where $\tilde{q}$ and $\phi$ are the relative transverse momentum and angle between the $c$ and the $\bar c$ in the rest frame of the pair. The charm quark mass is $m_c$, and $\mathcal{J}$ is the Jacobian of the variable transformation.

\emph{Non-relativistic QCD:} In NRQCD the direct production of a heavy quarkonium $H$ is computed by first evaluating the short distance coefficients, $\der \hat{\sigma}^{\kappa}$, for the production of a heavy quark pair in a given quantum state $\kappa = {}^{2S+1} L^{[C]}_{J} $,
where $S$, $L$, $J$, $C$ are the spin, orbital angular momentum, total angular momentum, and color state of the pair,
respectively. These coefficients are then weighted with non-perturbative long distance matrix elements $\langle \mathcal{O}^{H}_{\kappa}\rangle$ (LDMEs)\footnote{This is in contrast to the Color Evaporation Model where effectively all states are weighted by the same universal constant $F$.} and summed:
\begin{align}
    \der \sigma_H = \sum_{\kappa} \der \hat{\sigma}^{\kappa} \langle \mathcal{O}^{H}_{\kappa} \rangle \,.
\end{align}
Analogously, we express the direct yield for $J/\psi$ production in proton-nucleus collisions as
\begin{align}
\label{eq:cs}
    \frac{\der N_{J/\psi}(\bt)}{\der^2 \Pt \der Y} = \sum_{\kappa} \frac{\der N^{\kappa}_{c\bar c}(\bt)}{\der^2 \Pt \der Y} \langle \mathcal{O}^{J/\psi}_{\kappa}  \rangle \,.
\end{align}
Only four LDMEs contribute to $J/\psi$ production: three for the color octet state $\langle \mathcal{O}^{J/\psi}_{{}^{1} S^{[8]}_0} \rangle\!=\!0.089$~GeV$^3$, $\langle \mathcal{O}^{J/\psi}_{{}^{3} S^{[8]}_1} \rangle\!=\!0.0030$~GeV$^3$, and $\langle \mathcal{O}^{J/\psi}_{{}^{3} P^{[8]}_J} \rangle\!=\!0.0056$~GeV$^3$, and one for the color singlet: $\langle \mathcal{O}^{J/\psi}_{{}^{3} S^{[1]}_1} \rangle=1.16/(2N_c)$~GeV$^3$ \cite{Ma:2015sia}.

The differential $c\bar{c}$ yields for a given quantum state $\kappa$ are perturbatively computable  \cite{Kang:2013hta}:
\begin{align}
    &\frac{\der N^{\kappa}_{c\bar c}(\bt)}{\der^2 \Pt \der Y }=\frac{\alpha_s}{(2\pi)^9(N^2_c-1)} \int_{\ktone,\kt,\kt',\Rt}  \!\!\!\!\!\!\!\!\!\!\!\!\!\!\!\!\!\!\!\!\!\!\!\!\!\!\!\!\!\! \mathcal{H}^{\kappa}(\Pt-\k_{1,\perp},\k_{\perp},\kt')\notag \\ 
    & ~~~~~~~~~~\times \frac{\phi^{p}(x_{p},\ktone,\Rt)}{\ktone^2} \notag\\ &~~~~~~~~~~~\times\widetilde{\Xi}^{\kappa}(x_A;\Pt-\ktone,\kt,\kt';\Rt-\bt)   \,,
    \label{eq:cgc-ccbar-nqrcd}
\end{align}

The hard factors $\mathcal{H}^{\kappa}$ are given in \cite{Kang:2013hta}. The meaning of the $\ktone$, $\kt$ and $\Rt$ are the same as in Fig.\,\ref{fig:ccbarprod}. The variable $\kt'$ is the transverse momentum transfer to the quark in the conjugate amplitude \footnote{The momentum of the incoming gluon (from the proton) is approximately the same in the amplitude and conjugate amplitude. This results in a Dirac delta function $\delta^{(2)}(\ktone-\ktone')$; hence, the absence of $\ktone'$ in  Eq.\,\eqref{eq:cgc-ccbar-nqrcd} \cite{Blaizot:2004wu,Blaizot:2004wv}.}. 
The color correlator $\widetilde{\Xi}^{\kappa}$ depends on the color state of the $c\bar{c}$, octet $[8]$ or singlet $[1]$, which at
leading order in $N_c$ are given by
\begin{align}
    \widetilde{\Xi}^{[8]}(x;\lt,\kt,\kt';\Rt) &= (2\pi)^2 \delta^{(2)}(\kt-\kt') \notag\\ 
    & \hspace{-3.3cm}\times \widetilde{\mathcal{S}}_{\rm F}^A(x;\kt;\Rt)  \widetilde{\mathcal{S}}_{\rm F}^A(x;\lt-\kt;\R_\perp) + \mathcal{O}\left(1/N_c\right), \!\! \\
    \widetilde{\Xi}^{[1]}(x;\lt,\kt,\kt';\Rt) &= \widetilde{\mathcal{S}}_{\rm F}^A(x;\kt;\Rt)  \widetilde{\mathcal{S}}_{\rm F}^A(x;\kt';\Rt) \notag\\ 
    &\hspace{-2cm}\times \widetilde{\mathcal{S}}_{\rm F}^A(x;\lt-\kt-\kt',\R_\perp) + \mathcal{O}\left(1/N_c\right). \!\!
    \label{eq:color-singlet}
\end{align}
Observe that the color octet $\widetilde{\Xi}^{[8]}$ is proportional to $\delta^{(2)}(\kt-\kt')$, which implies that the momentum transfer to the quark (antiquark) is the same in the amplitude and conjugate amplitude, thus the yield for the octet production in NRQCD takes the same form as in Eq.\,\eqref{eq:cgc-ccbar}. Second, to obtain the color singlet $\widetilde{\Xi}^{[1]}$, we took the approximation proposed in \cite{Ma:2014mri} to simplify the quadrupole correlator as a product of dipoles resulting in Eq.\,\eqref{eq:color-singlet}.\footnote{This approximation also modifies the hard factor $\mathcal{H}^{{}^{3} S^{[1]}_1}$, which is given in \cite{Ma:2014mri}.} A systematic study of the range of validity of that approximation will be assessed in a separate publication. Third, while the color singlet is suppressed relative to the color octet, in NRQCD all states are weighted with different LDMEs; hence, both color states have to be kept. 

When also integrating over the $J/\psi$'s transverse momentum in Eq.\,\eqref{eq:cgc-ccbar-nqrcd}, for every event one needs to compute up to 10 dimensional integrals, which we do with Monte Carlo integration using a Vegas algorithm \cite{PETERLEPAGE1978192}.

\subsection{Two-point correlator}
 Next, we determine the two-point correlator (dipole) $S_{\rm{R}}$. We assume that the spatial and energy dependence of the dipole factorize, similar to the IPSat model \cite{Bartels:2002cj,Kowalski:2003hm}. We base our description on the impact parameter independent running coupling BK (rcBK) evolved dipole amplitudes, fitted to HERA data for proton structure functions \cite{Lappi:2013zma}.\footnote{We have also explored a simpler dipole, using the McLerran Venugopalan (MV) model \cite{McLerran:1993ni,McLerran:1993ka} and, with typical evolution speeds, have not found large differences compared to the rcBK evolved case for the studied observables.} To include a spatial dependence we first parametrize the rcBK evolved dipole amplitude in the fundamental representation as
\begin{align}
\label{eq:dipoleParam}
    S_{\mathrm{F}}\left(x; \frac{\rt}{2} ,  - \frac{\rt}{2} \right)
    = e^{-\frac{r^2_\perp Q_{s0}^2(x,\rt)}{4}\ln\left(\frac{1}{r_\perp\Lambda}+e\right)}\,,
\end{align}
where we used $\Lambda=0.2\,{\rm GeV}$ and
\begin{equation}
    Q_{s0}^{2}(x,\rt) = c_0 \left(\frac{0.01}{x}\right)^{0.3 \exp(-c_1 r_\perp)/(c_2 r_\perp)^{0.05}} \!\!\!\!
\end{equation} 
with $c_0= 0.104\,{\rm GeV}^2$, $c_1=0.16\,{\rm GeV}$ and $c_2=1\,{\rm GeV}$.
This is only an approximate representation of the full rcBK result, that reflects the most relevant differences to the MV form \cite{McLerran:1993ni,McLerran:1993ka} by incorporating the $r_\perp$ dependent $Q_s$ above. 
We model the impact parameter dependence of the dipole, by modifying the saturation scale:
\begin{align}\label{eq:A}
    Q_{sA}^{2}(x,\rt,\Rt) = T_{A}(\Rt) S_\perp Q_{s0}^{2}(x,\rt) \,,
\end{align}
where $S_\perp$ represents the transverse area of the proton in the $\Rt$-independent case \cite{Lappi:2013zma}.\footnote{$S_\perp$, which is constrained by matching to the experimentally determined $J/\psi$ multiplicity, is approximately 30\% smaller than that used in \cite{Lappi:2013zma}. The difference could be due to the impact parameter dependence we have introduced.}
This procedure yields the Fourier transform of the two-point correlator in the fundamental representation introduced in Eq.\,\eqref{eq:2pointWilson-coordinate}; in order to obtain the corresponding correlator in the adjoint representation we use the following relation
\begin{align}
    S_{\rm{Adj}}(x;\xt,\xt') = S_{\rm{F}}(x;\xt,\xt')^{\frac{N_c}{C_F}} \,.
\end{align}
The thickness function, $T_{A}(\Rt)$, encodes the spatial distribution of color charges in either projectile or target and can fluctuate on an event-by-event basis. 

The proton is described as composed of $N_q$ hot spots, whose centers, $\R_{\perp,i}$, are distributed in the transverse (to the beam line) plane according to a Gaussian distribution with width  $\sqrt{B_{\rm qc}}$. The transverse density profile of each hot spot is a Gaussian with width $\sqrt{B_q}$, i.e.,
\begin{equation}\label{eq:Tq}
    T_q(\R_\perp-\R_{\perp,i})=\xi_{Q_s^2} e^{-(\R_\perp-\R_{\perp,i})^2/(2 \xi_{B_q} B_q)}\,,
\end{equation}
and, consequently, the normalised proton thickness function is given by
\begin{equation}\label{eq:Tp}
    T_p(\R_\perp)=\frac{1}{2\pi N_q B_q }\displaystyle\sum_{i=1}^{N_q} T_q(\R_\perp-\R_{\perp,i})\,.
\end{equation}
In Eq.\,\eqref{eq:Tq}, we have introduced two scaling parameters, $\xi_{Q_s^2}$ and $\xi_{B_q}$, that fluctuate on an event-by-event basis according to log-normal distributions with widths $\sigma_{Q_s^2}$ and $\sigma_{B_q}$, respectively, and  have $\langle \xi_{Q_s^2}\rangle =\langle \xi_{B_q}\rangle =1$. The former allows the saturation scale normalization of each hot spot to fluctuate, while $\xi_{B_q}$ accounts for hot spot size fluctuations. Note that the normalization factor is always using the average $B_q$, without the factor of $\xi_{B_q}$.

This last fluctuation was not included in previous works \cite{Mantysaari:2016ykx}. To justify its inclusion, consider the following physical picture. Hot spots are the result of soft gluons radiated around a large-$x$ parton in the evolution. These hard partons are generated in the first splittings of the dipole cascade and the transverse separation of these branchings can in principle fluctuate\,\cite{Mueller:2010fi}. This can lead to hot spots of different sizes after evolution. It is beyond the scope of this work to provide a rigorous derivation of these size fluctuations. Here, we are interested in their phenomenological impact.
 
To model the nucleus geometry, we sample its nucleon positions according to a Woods-Saxon distribution with radius $R_A=6.62$~fm and diffuseness $a=0.546$, and add a fluctuating $T_p$ for each nucleon. 
 
We allow the hot spot width to grow with decreasing $x$ to approximate the growth of the proton observed in JIMWLK calculations \cite{Schlichting:2014ipa,Mantysaari:2018zdd}, and long known as Gribov diffusion \cite{Gribov:1973jg}. This effect is important to describe the center of mass energy dependence of the slope of $|t|$ spectra in diffractive $J/\psi$ production at HERA \cite{Mantysaari:2018zdd}, as well as the hollowness effect in $p+p$ collisions \cite{Albacete:2016pmp}. We employ a quadratic growth of $B_q$ following \cite{Schlichting:2014ipa} and \cite{Kovner:2001bh}.
Explicitly, we multiply $B_q$ (in both Eq.\,\eqref{eq:Tq} and \eqref{eq:Tp}) by the function $w(y) = 0.15 + 0.042*(\pm y - 4.6)^2$, where the plus is used for the right moving proton, and minus for the left moving nucleus. 

The parameters of our model are constrained with data on charged hadron and $J/\psi$ rapidity and transverse momentum distributions in $p+p$ and $p+Pb$ collisions, as well as the correlation between the charged hadron and $J/\psi$ multiplicities, which we focus on in this letter. Parameter values used are presented in Table~\ref{table:param}. 
The value for $B_{qc}$ is consistent with that obtained from fits to diffractive $J/\psi$ production at HERA \cite{Mantysaari:2016jaz}. The $B_{q}$ we employ at mid-rapidity is larger, but taking into account the growth of the proton with decreasing $x$ and the difference in $x$ values in $8.16\,{\rm TeV}$ collisions at LHC and those probed in the HERA data considered in \cite{Mantysaari:2016jaz}, the values are also consistent. In fact, using our $w(y)$ equation above, the size at a ten times larger $x$ should be $0.4\,{\rm GeV^{-2}}$, very close to the value found in \cite{Mantysaari:2016jaz}.

The impact parameter $\b_\perp$, the vector connecting the projectile and target centers, is sampled according to $P(b_\perp) \propto b_\perp$ (and uniformly distributed angle), along with the spatial dependence of the dipole amplitudes, allowing for a fully event-by-event study. Final results of the presented observables are averaged over many thousand events.

\begin{figure*}[htb]
    \centering
  \includegraphics[width=0.48\textwidth]{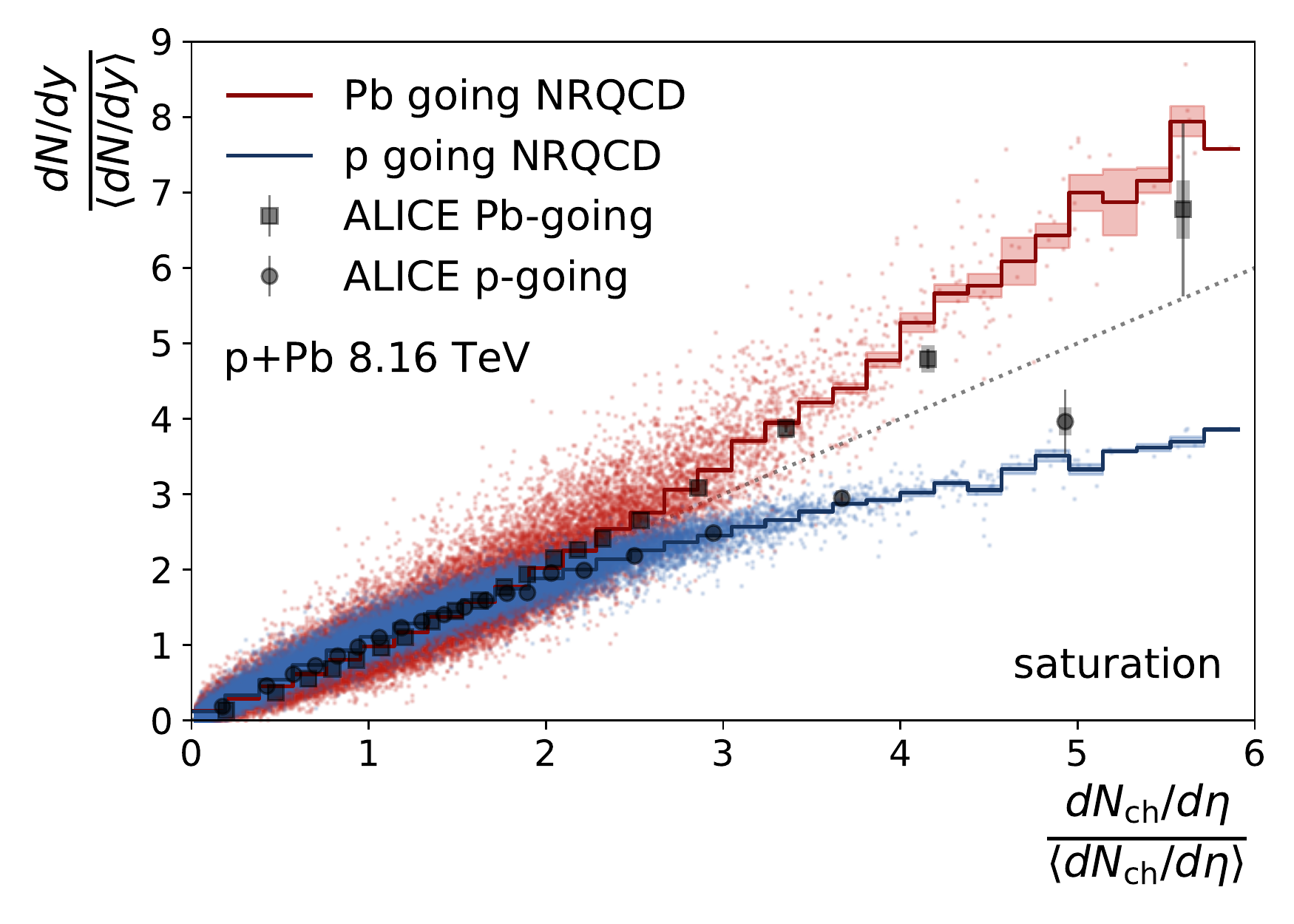}  \includegraphics[width=0.48\textwidth]{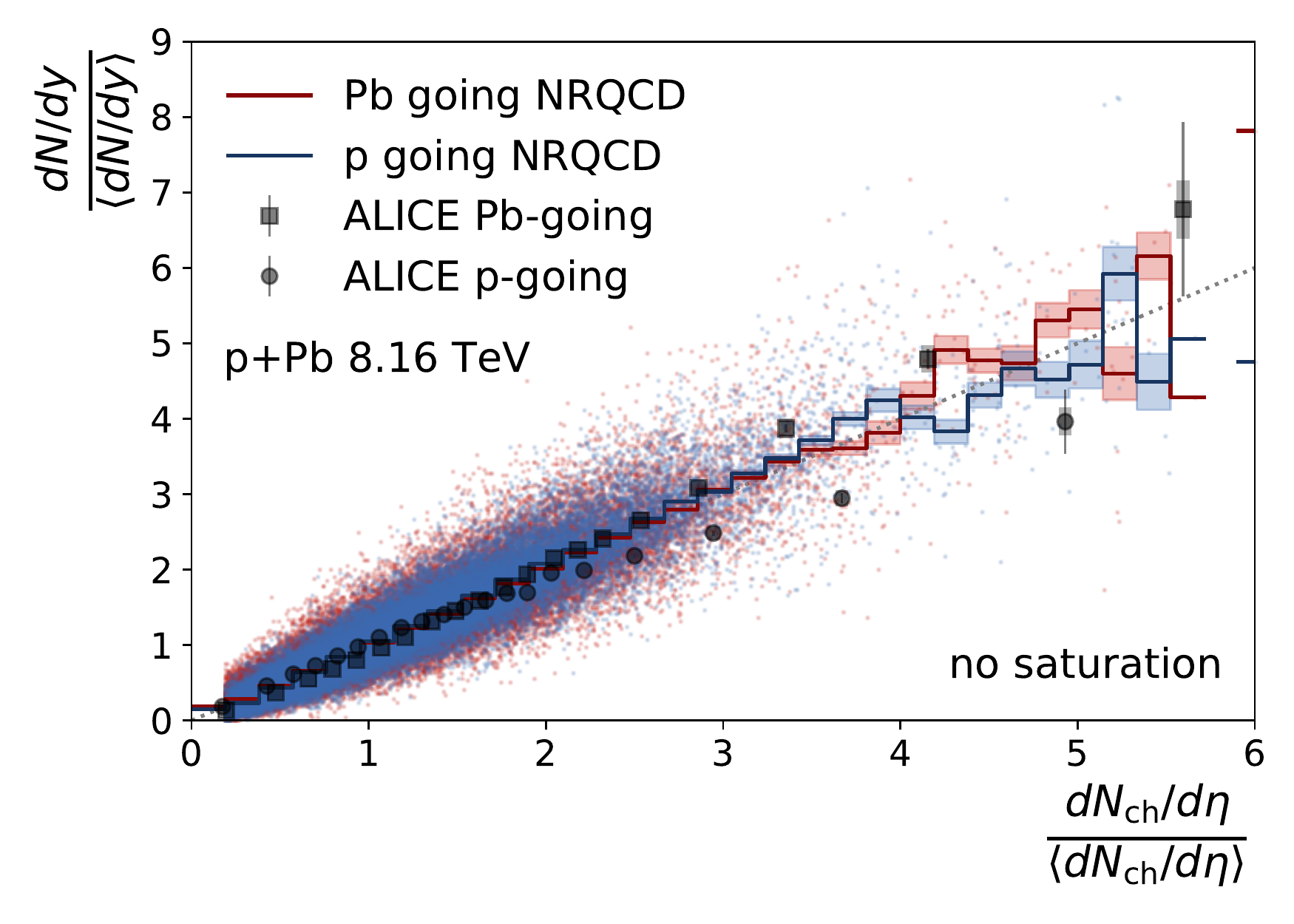}
     \vspace{-0.4cm}
     \caption{Correlation between normalized $J/\psi$ and charged hadron yields in $p+Pb$ collisions at 8.16 TeV for BK evolved dipoles and using NRQCD to describe the $J/\psi$ hadronization. Shown are results from individual events as scatter plots and the event average with statistical errors. Experimental data from ALICE \cite{ALICE:2020eji}. Left: Using values of $Q_s$ that lead to multiplicities in line with experimental results. Right: Artificially low $Q_s$ to mimic the situation of no, or much weaker, saturation effects.}
     \label{fig:correlator_scatter}
 \end{figure*}
 
\begin{table}[t]
\centering
\begin{tabular}{|l|l|}
\hline
Parameter   & Value \\
\hline
$N_q$           & 3\\
$B_{qc}$       & $3\,{\rm GeV}^{-2}$\\
$B_q$          & $1\,{\rm GeV}^{-2}$\\
$\sigma_{B_q}$ & 0.7\\
$\sigma_{Q_s^2}$ & 0.1\\
$S_\perp$   & $13\,{\rm mb}$\\
\hline
\end{tabular}
\begin{tabular}{|l|l|}
\hline
Parameter     & Value \\
\hline
$\alpha_s$     & 0.16\\
$m_{\rm IR}$   & $0.2\,{\rm GeV}$\\
$m_{J/\psi}$  & $3.1\,{\rm GeV}$\\
$m_{c}$       & $m_{J/\psi}/2$\\
$m_D$         & $1.87\,{\rm GeV}$ \\
~&~\\
\hline
\end{tabular}
\caption{\label{table:param}Standard set of parameters used. To constrain the normalization of multiplicities we use the experimentally determined inelastic cross section $\sigma_{\rm inel}= 2161\,{\rm mb}$ for $\sqrt{s}=8.16\,{\rm TeV}$ $p+Pb$ collisions (extrapolated from result at $5.02\,{\rm TeV}$ \cite{Baur:2015rsb}), and $\sigma_{\rm inel}= 60\,{\rm mb}$ for $7\,{\rm TeV}$ $p+p$ collisions \cite{CMS:2012gek}.}
\end{table}

\begin{table}[t]
	\centering\small
	\begin{tabular}{|c|c|c|c|}
		\hline
		System & $\sqrt{s_{{\rm NN}}}$ [TeV] & $J/\psi$ rapidity range & $J/\psi$ rapidity used\\
		\hline
		\multirow{2}{*} {$p+p$} & \multirow{2}{*} {$7$}  & \multirow{1}{*} {$|y|<0.9$} & 0\\
		\cline{3-4}
	    & & 	{$2.5<y<4$} & 3.03\\
		\hline
		\multirow{2}{*}[-1.7em]{$p+Pb$} & \multirow{2}{*}[-0.7em]{$5.02$} & \multirow{1}{*}{$-4.46<y<-2.96$} & -3.52 \\ \cline{3-4}
		& & \multirow{1}{*}{$-1.37<y<0.43$} & -0.47\\ \cline{3-4}
		& & \multirow{1}{*}{$2.03<y<3.53$} & 2.62\\ 	\cline{2-4}
		& \multirow{2}{*}{$8.16$} & \multirow{1}{*}{$-4.46<y<-2.96$} & -3.52 \\ \cline{3-4}
		& & \multirow{1}{*}{$2.03<y<3.53$} & 2.62 \\
		\hline
	\end{tabular}
	\caption{\label{table:data} Kinematic configurations of the experimental data presented in~\cite{ALICE:2012pet,ALICE:2017wet,Acharya:2018egz}, and rapidity values used in our calculations. $p$-going: $y>0$, $Pb$-going: $y<0$.}
\end{table}
\section{Results}

 \begin{figure}[ht]
    \centering
    \includegraphics[width=\columnwidth]{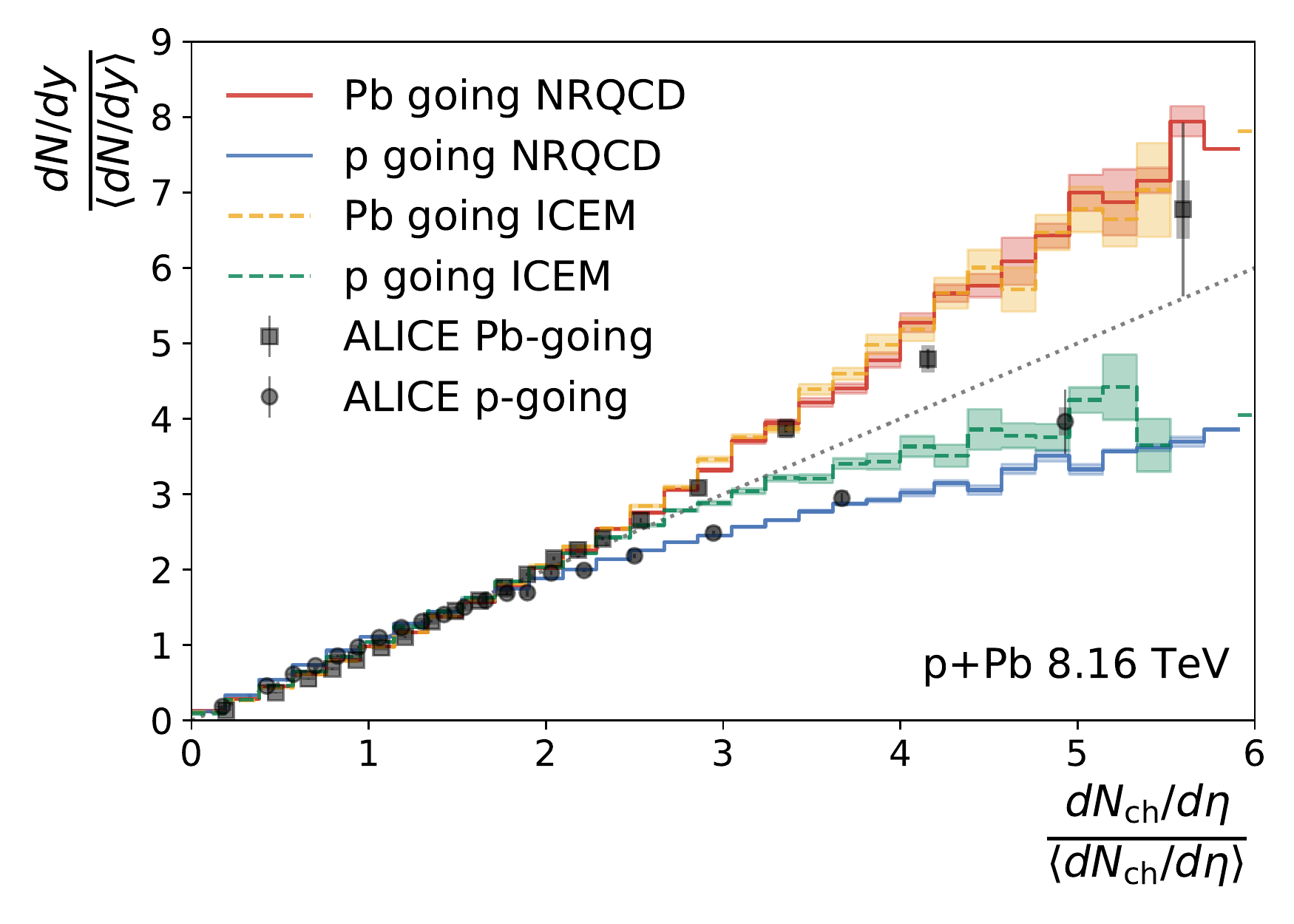}
    \vspace{-0.8cm}
     \caption{Correlation between normalized $J/\psi$ and charged hadron yields in 8.16 TeV $p+Pb$ collisions using either NRQCD (solid) or ICEM (dashed) for the $J/\psi$ hadronization.}
     \label{fig:correlator_std}
 \end{figure}

We compute the $J/\psi$ multiplicity, normalized by its mean value, and present its correlation with the similarly normalized charged hadron multiplicity in the left panel of Fig.\,\ref{fig:correlator_scatter}.
Charged hadrons are always measured at mid-rapidity $|\eta|<1$ (we choose $\eta=0$ in the calculation), while the vector mesons are computed in different rapidity windows, shown in Table~\ref{table:data}. For simplicity, we do not integrate over the rapidity interval in the theoretical calculation, but use a central value (weighted by the multiplicity), also shown in Table~\ref{table:data}. Events with $dN_{\rm ch}/d\eta<0.5$ are rejected, which is approximately equivalent to requiring at least one charged hadron within two units of rapidity.
 
We find a very good agreement between our CGC based model and the experimental data at negative ($Pb$-going) and positive ($p$-going) rapidities. We show the results for individual events as a scatter plot to highlight the width of the distributions. It would be highly beneficial for further constraining theoretical models if a similar measure of fluctuations could be accessed experimentally. 

The correlator deviates most significantly from the linear trend in rare, high multiplicity events which, in our framework, are generated by upward fluctuations of the saturation scales or interaction region, arising from geometry fluctuations, including the hot spot size fluctuations, or the log-normal fluctuations of the saturation scale itself. 

The observed rapidity dependence of the correlator is naturally explained in the saturation picture. Hadrons are measured at mid-rapidity, where the values of $x$ in projectile and target are of the order of $\langle p_T \rangle/\sqrt{s} \approx 10^{-4}$. Therefore, even in the proton, typical $Q_s$ values are approximately $0.6\,{\rm GeV}$, such that hadron production, known to scale as $N_{\rm ch}\propto{\rm min}(Q_{sp},Q_{sA})$ \cite{Dumitru:2001ux}, is expected to be sensitive to saturation. In the backward direction ($Pb$-going), the mass of the $J/\psi$ is typically larger than the saturation scales in both the proton and the nucleus. Therefore, a stronger than linear increase, caused by saturation of the hadrons but not of the $J/\psi$, is observed in the data. In contrast, in the $p$-going direction, $Q_s$ values in the nucleus can reach values of the order of the $J/\psi$ mass and larger, and the $J/\psi$ multiplicity is strongly suppressed when correlated with the event activity at mid-rapidity.

To emphasize the importance of these saturation effects, we mimic a scenario without saturation by artificially decreasing the value of $S_\perp$ in Eq.\,\eqref{eq:A} to $1\,{\rm mb}$. This renders $Q_s$ values entering the calculation smaller than any relevant momentum scale in most cases. Of course this rescaling leads to a much too low normalization of the charged hadron and $J/\psi$ yields, but we can still study the correlation of the normalized multiplicities. The result is shown in the right panel of Fig.\,\ref{fig:correlator_scatter}, where we find no significant difference between the $p$-going and $Pb$-going direction, with both results being much closer to the diagonal, as one would expect if saturation effects do not play a role. The agreement with data is notably worse. 

In the following, we evaluate the robustness of the result presented in the left panel of Fig.~\ref{fig:correlator_scatter}, which has the correct multiplicity normalization and saturation effects, against variations in the model parameters.

In Fig.~\ref{fig:correlator_std}, we study the sensitivity to the production model of the $J/\psi$, by comparing results obtained from NRQCD with those of the ICEM.
While the backward direction is rather insensitive to the choice of vector meson production model, the result in the $p$-going direction moves up when going from NRQCD to ICEM, overestimating the data for most points beyond $(dN_{\rm ch}/d\eta)/\langle dN_{\rm ch}/d\eta \rangle >2$. This result is closer to previous CGC based calculations using the ICEM \cite{Ma:2021epm}, but we still see more suppression. Because of the non-perturbative nature of the hadronization prescription, it is hard to interpret the origin of the difference, but this exercise provides us with an estimate of the systematic uncertainty originating from the description of the $J/\psi$ hadronization. 

\begin{figure}[tb]
     \centering
     \includegraphics[width=\columnwidth]{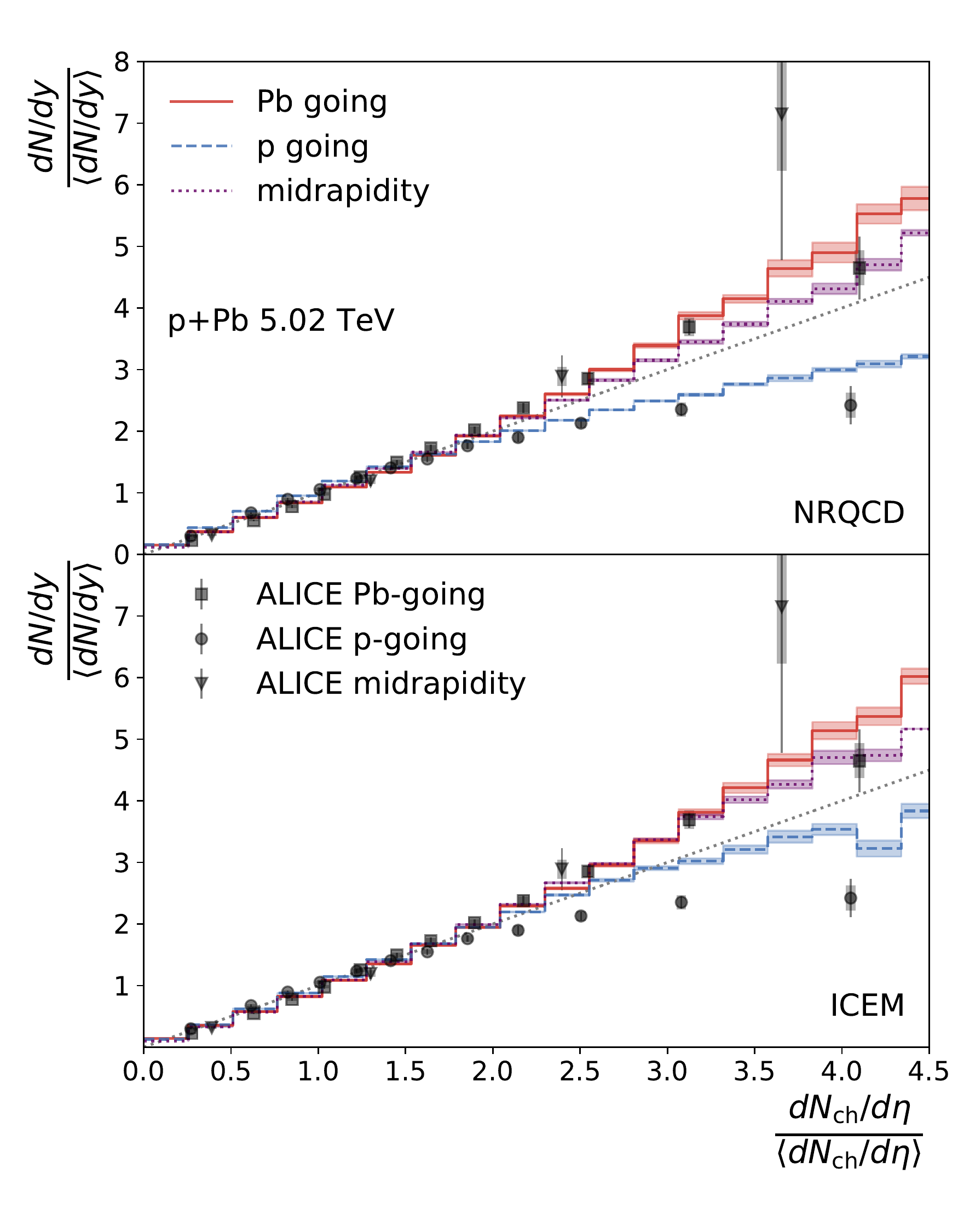} \vspace{-1cm}
     \caption{Correlation between normalized $J/\psi$ and charged hadron yields in 5.02 TeV $p+Pb$ collisions, using either NRQCD (top) or ICEM (bottom) for the $J/\psi$ hadronization. Experimental data from the ALICE Collaboration \cite{ALICE:2017wet}.}
     \label{fig:correlator_5020}
 \end{figure}

A similar dependence on the hadronization prescription of the $J/\psi$ is observed for $5.02\,{\rm TeV}$ collisions, as shown in  Fig.~\ref{fig:correlator_5020}. As for this energy there also exist measurements of $J/\psi$ at mid-rapidity \cite{ALICE:2017wet}, we include this rapidity bin in the calculation. The figure shows results using NRQCD and ICEM in the upper and lower panels, respectively. Results for $p$-going and $Pb$-going directions are similar to the case of $8.16\,{\rm TeV}$ collisions. The correlator changes monotonically with rapidity, most clearly for the NRQCD case, while the experimentally measured correlator for the $J/\psi$ at mid-rapidity increases more quickly than that with $J/\psi$ observed in the $Pb$-going direction. In the ICEM case, the mid-rapidity result is much closer to the $Pb$-going result, and for some multiplicities the ordering is like that in the experimental data. Nevertheless, reproducing the stronger rise at mid-rapidity compared to the $Pb$-going direction is very challenging in the CGC framework, for any hadronization prescription. 

\begin{figure}[tb]
     \centering
     \includegraphics[width=\columnwidth]{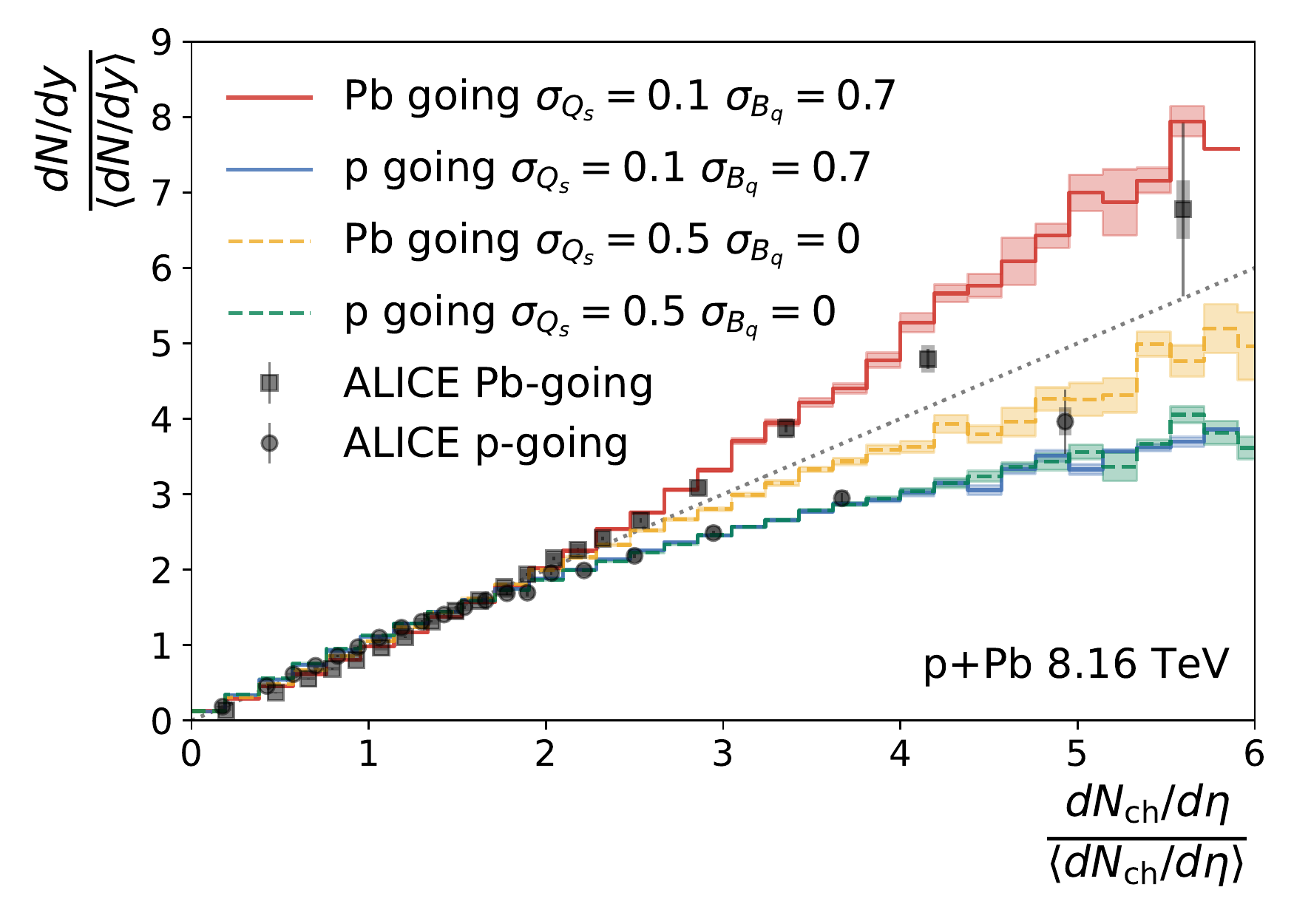}
     \caption{Correlation between normalized $J/\psi$ and charged hadron yields in  8.16 TeV $p+Pb$ collisions, using NRQCD for the $J/\psi$ hadronization. Comparison of simulations with size fluctuations (solid) to ones with only (increased) $Q_s$ normalization fluctuations (dashed).}
     \label{fig:correlator_w05}
 \end{figure}

To demonstrate that besides its sensitivity to saturation effects, this observable is also sensitive to the details of the geometrical fluctuations, we compare our standard scenario to one where the $Q_s$ normalization strongly fluctuates, while the hot spot size varies mildly, in Fig.\,\ref{fig:correlator_w05}. In our standard scenario, large hadron multiplicities can be generated by either (a) upwards fluctuations of the saturation scale or (b) fluctuations to larger protons, without going too deeply into the saturated regime. Contrarily, in the situation with large $Q_s$ normalization fluctuations, large hadron multiplicities can only be reached via fluctuations to very dense hot spots (in both proton and nucleus). Thus, saturation effects on the $J/\psi$ become much larger than in our standard scenario, and the correlator moves below the diagonal in the $Pb$-going direction. 
The $p$-going direction is unmodified when changing the fluctuations. In this case, the saturation scale is much larger, and details of its fluctuations seem to be less important. 

Clearly, the large normalization fluctuations are incompatible with the experimental data. Consequently, the study of this observable provides additional constraints on the details of subnucleonic fluctuations, that HERA data on diffractive $J/\psi$ data could not restrict \cite{Mantysaari:2016ykx}. It remains to be checked whether the addition of size fluctuations and reduced normalization fluctuations are also compatible with the diffractive HERA data. In any case, we have demonstrated that including the data studied here, together with $e+p$ and other data, in a global Bayesian analysis is likely to be very beneficial for constraining models of subnucleonic fluctuations.

\begin{figure}[tb]
     \centering
     \includegraphics[width=\columnwidth]{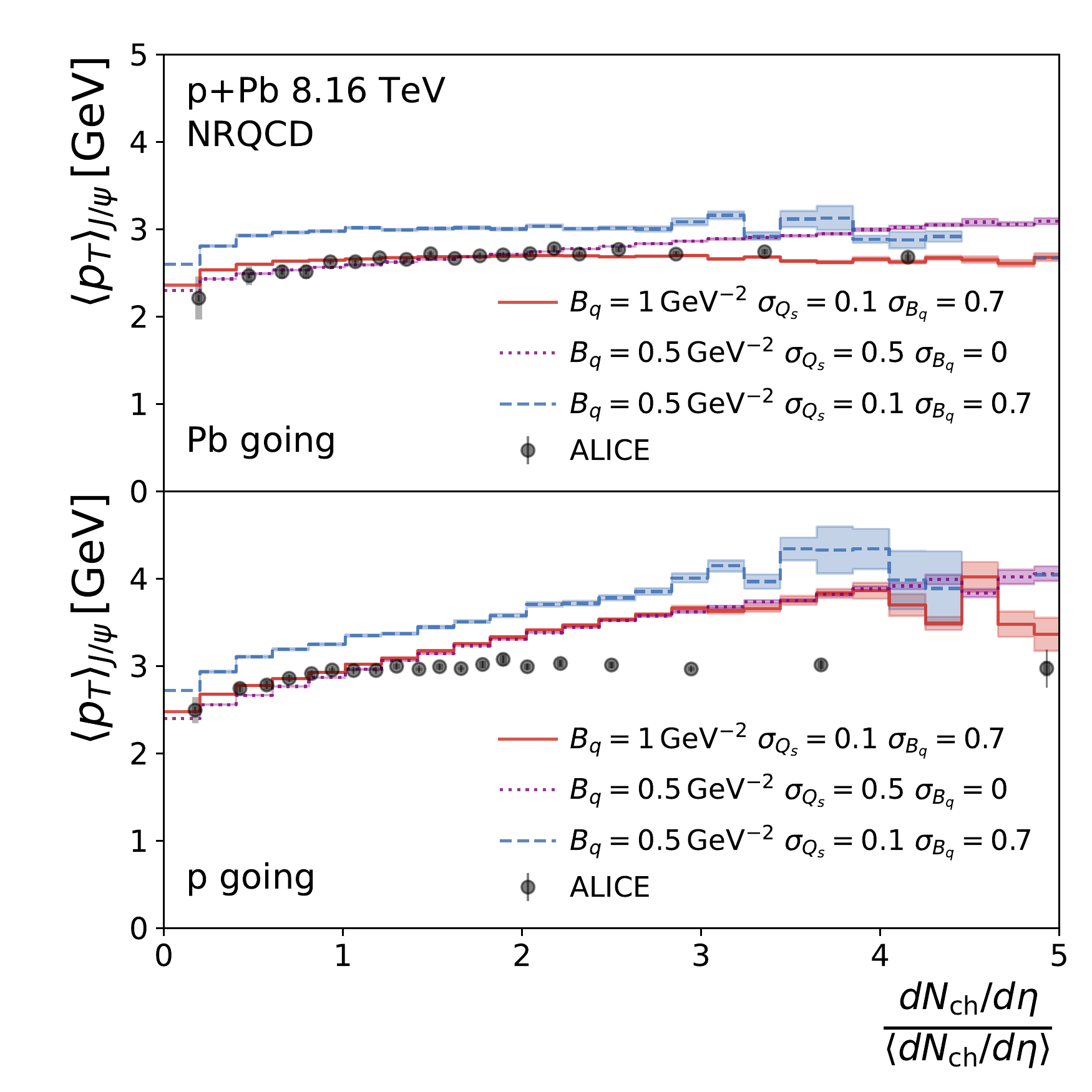}
     \caption{Mean transverse momentum of the $J/\psi$ as a function of $\der N_{\rm ch}/\der \eta$ in the $Pb$- (upper) and $p$-going (lower) directions, compared to experimental data from ALICE \cite{ALICE:2020eji}. Comparing the standard parameter set (solid) to the case using smaller hot spots (dashed) and the case without size fluctuations, but larger $Q_s$ normalization fluctuations (dotted).}
     \label{fig:meanpT_JPsi}
 \end{figure}

Another important observable that is highly sensitive to geometrical fluctuations and saturation is the $J/\psi$ mean transverse momentum $\langle p_T\rangle$. In Fig.\,\ref{fig:meanpT_JPsi}, we show both $Pb$-going and $p$-going $J/\psi$ $\langle p_T \rangle$, compared to experimental data from the ALICE Collaboration \cite{ALICE:2020eji}. For our standard value $B_q=1\,{\rm GeV}^{-2}$, the $Pb$-going measurement is very well described, while our $p$-going result agrees with the data for up to approximately 1.5 times the average multiplicity, but above that shows a stronger multiplicity dependence than the surprisingly flat experimental data.

In the CGC framework the $\langle p_T\rangle$ of produced particles is expected to be driven by the saturation scale. In fact, in the $p$-going direction, we observe that our $\langle p_T\rangle$ result monotonously increases with hadron multiplicity, i.e., the larger the saturation scale, the larger $\langle p_T\rangle$ of the $J/\psi$. This trend is not observed in the $Pb$-going direction, because there typically $Q_s \ll m_{J/\psi}$, making the mass the dominant scale for $\langle p_T\rangle$. 

When turning off hot spot size fluctuations and increasing $Q_s$ normalization fluctuations (c.f.~Fig.\,\ref{fig:correlator_w05}), we see a rising $\langle p_T\rangle$ as a function of multiplicity in the $Pb$-going direction. In this case, large multiplicities are mainly reached by fluctuations to large $Q_s$ values (and not to larger interaction areas). Thus, the correlation between $\langle p_T \rangle$ and multiplicity becomes stronger. However, in the $p$-going direction, we see little difference between the two scenarios with different fluctuations. As the $p$-going direction corresponds to the deep saturation regime of the nucleus, it is possible that $Q_s$ fluctuations remain dominant, even when including size fluctuations.

For a smaller average value of $B_q=0.5\,{\rm GeV}^{-2}$, the $\langle p_T\rangle$ is larger for all multiplicities and both rapidities. This is a consequence of having a larger normalization for the smaller hot spots (see Eq.\,\ref{eq:Tp}), resulting in significantly larger $Q_s$ values, which increases the $\langle p_T\rangle$. This observable shows an enhanced sensitivity to the choice of $B_q$ as compared to the multiplicity correlators studied above.

\begin{figure}[tb]
     \centering
     \includegraphics[width=\columnwidth]{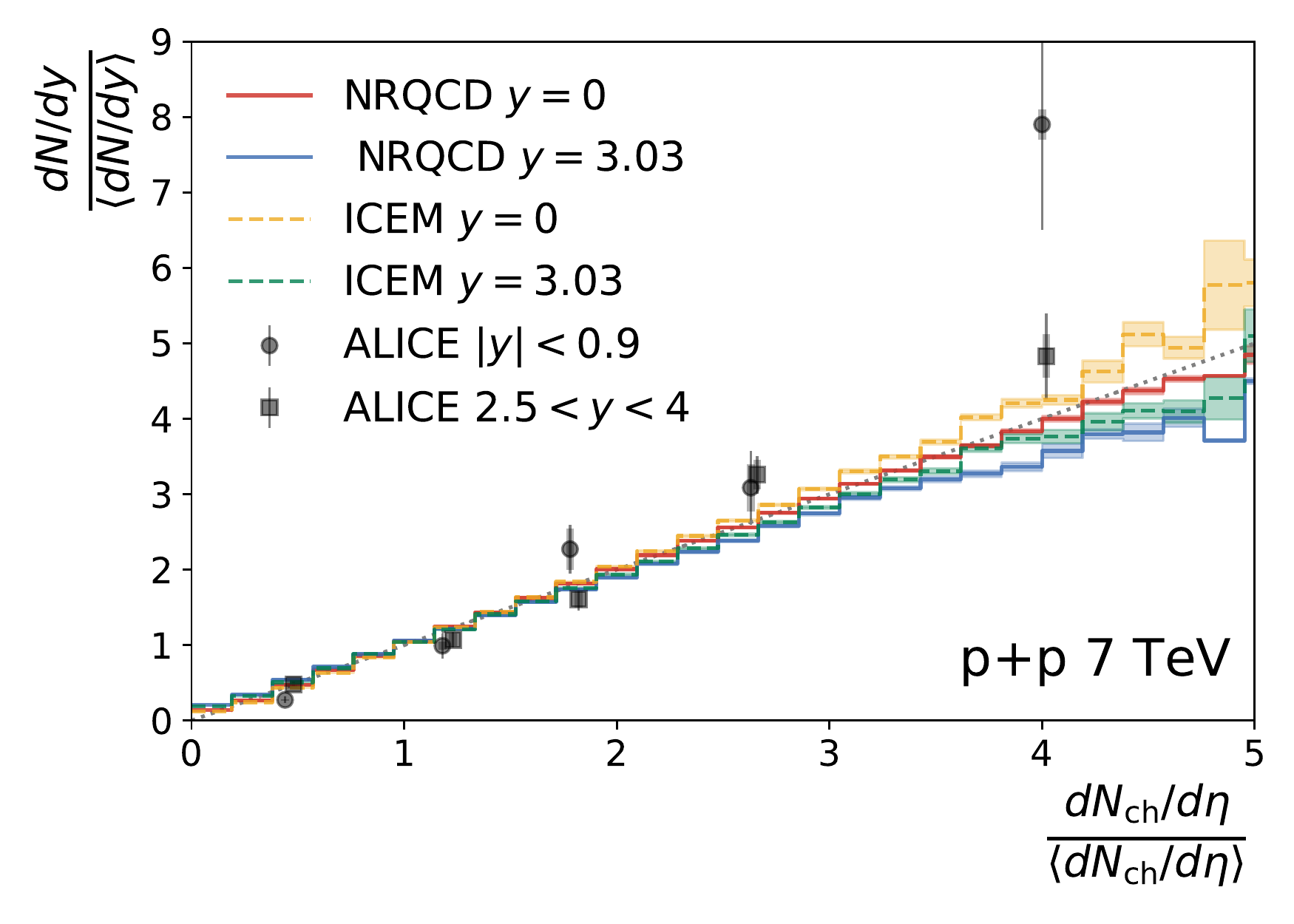}
     \caption{Correlation between normalized $J/\psi$ and charged hadron yields in $p+p$ collisions at 7 TeV, using either NRQCD (solid) or ICEM (dashed) to describe the $J/\psi$ hadronization. Experimental data from the ALICE Collaboration \cite{ALICE:2012pet}. }
     \label{fig:correlator_pp}
 \end{figure}

Finally, we show the multiplicity dependent $J/\psi$ production in $p+p$ collisions in Fig.~\ref{fig:correlator_pp}. The data shows that when hadrons and vector mesons are measured at mid-rapidity, the correlator grows faster than in the situation in which the $J/\psi$ is measured at forward rapidities. In a saturation picture, the forward data suggests saturation in one of the protons. This trend is captured by our theoretical model for both hadronization models. However, the model fails to quantitatively describe the experimental data from the ALICE Collaboration \cite{ALICE:2012pet}. The rise of the $J/\psi$ yield with charged hadron multiplicity is much weaker than experimentally observed.
In a previous CGC calculation \cite{Ma:2018bax}, which does not include any fluctuations, a stronger rise of the $J/\psi$ multiplicity with event activity was observed. Differences to such a model are expected, as in our case large charged hadron multiplicities can be reached from a variety of different fluctuations in either proton, whereas in \cite{Ma:2018bax} they are achieved by varying both protons' average $Q_s$ simultaneously. Nevertheless, the strong underestimation of the experimental data in $p+p$ and also for midrapidity $p+Pb$ (Fig.\,\ref{fig:correlator_5020})
in our model, hints that some physics is being missed.

\section{Conclusions}
We computed forward and backward $J/\psi$ production in $p+Pb$ collisions at LHC energies and its correlation with the mid-rapidity charged hadron multiplicity. We used the CGC framework to compute gluon and $c\bar{c}$ pair cross sections, employing rcBK evolved dipole amplitudes, KKP fragmentation of gluons to hadrons, and either NRQCD or the ICEM to describe the $J/\psi$ hadronization. The dependence of the $J/\psi$ multiplicity at both rapidities on the event activity was found to depend strongly on the presence of saturation effects, making this observable a sensitive probe of gluon saturation. Agreement with the experimental data required the presence of saturation effects.

The parameter values used for the subnucleonic structure are consistent with those obtained for diffractive $J/\psi$ production in $e+p$ collisions measured at HERA, when considering the $x$-dependence of the proton size. In addition, the observable studied here is particularly sensitive to the role of density vs.~size fluctuations, which highlights its potential to better constrain the fluctuating subnucleonic structure and the details of particle production. 

The faster than linear rise of the mid-rapidity $J/\psi$ yield with charged hadron multiplicity is weaker than in the experimental data, but it was found to be sensitive to the $J/\psi$ hadronization model, as is the $J/\psi$ production in the proton going (forward) direction.
In $p+p$ collisions, we see a significantly weaker increase of the $J/\psi$ yield with multiplicity than in the experimental data, with a similar dependence on the hadronization model.

For $p+Pb$ collisions, the $J/\psi$ mean transverse momentum as a function of charged hadron multiplicity in the $Pb$-going direction is well described using NRQCD, while the result in the $p$-going direction overestimates the data at high multiplicities. The mean-$p_T$ is also sensitive to the hot spot size, providing an additional constraint on that parameter. 

A natural continuation of this work would be to understand the systematics of our results for different heavy quark bound states such as $\Upsilon$~\cite{CMS:2013jsu} or $D$~\cite{ALICE:2015ikl}. In the former, we expect Sudakov resummation to become important \cite{Watanabe:2015yca}.
It will also be interesting to perform an event-by-event calculation using the dipole and quadrupole operators from JIMWLK evolved Wilson lines, which will naturally include the growth of the proton and modification of the geometry towards small-$x$. 

The inclusion of final state effects is also desirable. They have the potential to modify in particular the charged hadron transverse momentum (not shown), which in our calculation is well described at the lowest multiplicities, but underestimated at large multiplicities. It will also be interesting to see to what degree the $J/\psi$ is affected by final state effects.

Finally, this study suggests a strong potential for a combined global analysis of HERA (and future EIC) and LHC data to constrain (sub-) nucleon structure at high energies.

\section*{Acknowledgements}
We thank Javier Albacete, Zhongbo Kang, Yan-Qing Ma, Heikki M\"antysaari, Astrid Morreale, S\"oren Schlichting, Prithwish Tribedy, and Raju Venugopalan for useful discussions. We thank Raju Venugopalan and Kazuhiro Watanabe for helpful comments on the manuscript.
We thank Heikki M\"antysaari for providing the data for the rcBK evolved dipole amplitude. F.S. is supported by the National Science Foundation under grant No. PHY-1945471. B.P.S. is supported by the U.S. Department of Energy under Contract No. DE-SC0012704.
A.S.O.’s work was supported by the European Research Council (ERC) under the European Union’s Horizon 2020 research and innovation programme (grant agreement No. 788223, PanScales).
This research used resources of the National Energy Research Scientific Computing Center, which is supported by the Office of Science of the U.S. Department of Energy under Contract No. DE-AC02-05CH11231. 

\bibliography{references_jpsi}{}
\bibliographystyle{apsrev4-1}

\newpage

\end{document}